\documentclass[a4paper]{jpconf}
\usepackage{graphicx}
\begin{document}

\title{ Neutrinos: Windows to New Physics}

\author{R. D. Peccei}
\address{Department of Physics and Astronomy\\
University of California at Los Angeles\\
Los Angeles, California, 90095}

\ead{peccei@physics.ucla.edu}

\begin{abstract}

After briefly reviewing how the symmetries of the Standard Model (SM) are affected by neutrino masses and mixings, I discuss how these parameters may arise from GUTs and how patterns in the neutrino sector may reflect some underlying family symmetry. Leptogenesis provides a nice example of how different physical phenomena may be connected to the same neutrino window of physics beyond the SM. I end with some comments on the LSND signal and briefly discuss the idea that neutrinos have environment dependent masses.

\end{abstract}

\section{Symmetries of the Standard Model}

The Standard Model (SM) is invariant under local transformations of an
$SU(3)\times SU(2)\times U(1)$ group, which is spontaneously broken to
the subgroup $SU(3) \times U(1)_{em}$. However, in addition, at the
Lagrangian level the SM is invariant under four global $U(1)$
transformations, associated with baryon number $B$ and the individual
lepton numbers of each generation of leptons $L_i$. It is convenient to
group these symmetries into the set 
$U(1)_{\rm{B+L}}$; $U(1)_{\rm{B-L}}$; $U(1)_{\rm{L_i-L_j}}$ 
$[\rm{i,j=(1,2,3)}]$,
 where $L$ is the total lepton number,
$L=\Sigma_i L_i $. The last three symmetries are exact, but $U(1)_{\rm
{B+L}}$ has an electroweak anomaly, \cite{t'H} so it is not a symmetry
at the quantum level.  

The existence of neutrino masses and mixings, inferred from neutrino
oscillation experiments, alters these symmetry patterns and, hence,
provides evidence for physics beyond the SM. Neutrino mixings imply that
individual lepton number is not conserved. In effect, because of the
observed mixings one loses the two SM global symmetries associated with
$ U(1)_{\rm{L_i-L_j}}$. The existence of neutrino masses also affects
the symmetry structure of the SM, indicating the presence of new
interactions which most likely violate $U(1)_{\rm{B-L}}$.

\section{Disquisitions on Neutrino Masses}

Since neutrinos have zero charge, besides the usual Dirac
(particle-antiparticle) mass neutrinos can have also a Majorana
(particle-particle) mass. Using $\nu^c=C\nu^{\dagger}$, the most general mass term for neutrinos
has the structure:
\begin{equation}
L_{\rm{mass}}=-\frac{1}{2}
(\overline{\nu^c_L},\overline{\nu_R})\left[\begin{array}{cc}m_T&m_D^T\\ m_D&m_S
\end{array}\right]\left(\begin{array}{c}\nu_L\\ \nu^c_R
\end{array}\right) +h. c.
\end{equation}

It is clear from the above that $m_T$ and
$ m_S$ are Majorana mass terms, while $ m_D$ is a Dirac mass term. These
masses break the SM symmetries differently, since
$m_T$  is an $ SU(2)$ triplet  with (B-L)-charge
$Q_{\rm{B-L}} = -2$; $m_D$  is an $ SU(2)$ doublet  with (B-L)-charge $Q_{\rm{B-L}}
= 0$; while $m_S$  is an $ SU(2)$ singlet  with (B-L)-charge $Q_{\rm{B-L}}
= 2$.
Because of these different transformation properties, each of these mass terms is connected to different physics.
What combination of $m_T$, $m_S$, and $m_D$ determines the tiny
($m_{\nu}< 1$ eV) neutrino masses observed is a function of what this
underlying physics is. Below, I discuss each of these mass terms in
turn.

Perhaps the most conservative stance to adopt is to admit the existence of a right-handed neutrino $
\nu_R$, but assume that (B-L) remains a symmetry of nature. Then the
only neutrino mass term is $m_D$, which arises from the Yukawa coupling $\Gamma_{\nu}$
of the left-handed lepton doublet to $\nu_R$ and the Higgs field $\Phi$.
Because the observed neutrino masses are so small, one must
then understand what physics gives
$ m_D = \Gamma_{\nu}<\Phi> ~  <<~  m_{\ell} =\Gamma_{\ell} <\Phi>.$

One interesting possibility is to invoke extra dimensions. \cite{ADD} I
want to illustrate this idea by considering a  $d=5$ theory compactified
to  a space of radius $L$. In this theory, before compactification, the
Einstein action in 5-dimensions
 is characterized by a Planck mass $M^*$. After compactification                 the 4-dimensional Planck mass $M_P$ is related to its 5-dimensional
counterpart through the relation $ M_P^ 2 = M^{*3} L$. 
In such a 5-dimensional theory, it is natural to assume that while the
SM particles live on  the $D3$ brane, the right-handed neutrinos live in
the 5-dimensional bulk. Then, the neutrino action  involving $\nu_R$
reads
\begin{equation}
S=M^*\int d^4x \int dy
\sqrt{-g_5}\overline{\nu_R}\gamma^{\mu}i\partial_{\mu}\nu_R +\int d^4x
\Gamma_{\nu}^5\overline{\nu_R}\Phi L_L +h. c. ~.
\end{equation}
The first term in this action, after compactification becomes
\begin{equation}
M^*\int d^4x \int dy
\sqrt{-g_5}\overline{\nu_R}\gamma^{\mu}i\partial_{\mu}\nu_R \to M^*L \int
d^4x \sqrt{-g_4}\overline{\nu_R}\gamma^{\mu}i\partial_{\mu}\nu_R.
\end{equation}
Thus, to get the correct kinetic energy for the right-handed neutrinos,
one must rescale the $\nu_R$ field by a  factor $(M^*L)^{-1/2} \equiv
M^*/ M_P$.  This rescaling, effectively replaces the  Yukawa coupling
$\Gamma_{\nu}^5$, which one presumes is of the same order of
magnitude as $\Gamma_{\ell}$, by
$\Gamma_{\nu}\equiv \frac{M^*}{ M_P}\Gamma_{\nu}^5 $.
Hence, in these theories, one gets naturally a large reduction for the
right- handed neutrino Yukawa couplings.

There are many issues, however,
which remain open. For instance, why should (B-L) be conserved, given
that a right-handed neutrino exists. Given that $m_S$ breaks no
symmetries, besides (B-L), why is this mass term not permitted in the
theory? In addition, one must worry about the tower of Kaluza-Klein
states associated with the fact that $\nu_R$ lives in the bulk. What is
the effect of this tower of states and are these states dangerous?  This
suggests, that perhaps a better alternative is to have no $\nu_R$ states
in the theory at all.

If there are no right-handed neutrinos, neutrino masses can still arise from a
triplet mass. Because these masses break $SU(2)$, $m_T$ is proportional
to the VEV of a Higgs field which transforms as an $SU(2)$ triplet. This field is
either elementary, $\vec{T}$, or composite, $\vec{T}_{\rm{eff}} \equiv
\Phi^T C \vec{\tau}\Phi /\Lambda $, where $\Lambda $ is a new mass scale.
In the first case, small neutrino masses arise either because the
triplet Yukawa coupling is very small or because $<\vec{T}>~~<<~~ <\Phi> =
v_F \sim 250$ GeV. Because neither possibility is easily explainable, the
second alternative is much more natural. \cite {Weinberg} It ascribes
the smallness of neutrino masses to the presence of the large scale
$\Lambda$, associated with (B-L)-breaking, entering in the composite
operator $\vec{T}_{\rm{eff}}$. In this case, neutrino masses are given by the seesaw
formula
\begin{equation}
m_{\nu}= m_T \sim\frac{v_F^2}{\Lambda}.
\end{equation}
 and their presence reflects physics well beyond the  electroweak scale, since $\Lambda >>v_F$.

One can arrive at similar seesaw formula from singlet breaking,
\cite{seesaw} provided $m_S~ >>~ m_D$. Because $m_S$ carries no SM
quantum numbers, its scale is unconnected to $v_F$  and reflects
directly the scale of $U(1)_{\rm{B-L}}$ breaking.
If a hierarchy exists, $m_S~ >>~ m_D$, then the neutrino mass matrix
\begin{equation}
M= \left[\begin{array}{cc}0&m_D^T\\ m_D&m_S \end{array}\right]
 \end{equation}
has both large,  $M_N = m_S$ and small eigenvalues $M_{\nu}= - m_D^T
(m_S)^{-1} m_D$.

Let me summarize the principal lessons one has learned from the
existence of neutrino masses and mixings. They are two-fold:

\noindent i.) There are additional interactions beyond the SM, which
break all the global symmetries of the standard model.

\noindent ii.) Most likely, the small neutrino masses observed are
related to the large scale where $U(1)_{\rm{B-L}}$ breaks down.

\noindent In relation to the second point, it is not really possible to
tease apart whether $M_{\nu}$ arises from a triplet composite operator
[$M_{\nu}=m_T \sim <\vec{T}_{\rm{eff}}>\sim  v_F^2/\Lambda$], or large
right-handed neutrino masses [$M_{\nu}= - m_D^T (m_S)^{-1} m_D$], or a
combination of both mechanisms [$ M_{\nu}= m_T - m_D^T (m_S)^{-1} m_D$].

\section{Recondite Physics}

Looking beyond the SM for hints to the origins of neutrino masses and
mixing, there are three fruitful avenues to follow:

\noindent i.) A top-down approach, where one looks for possible new
symmetries and symmetry breakings associated with giving neutrinos
masses.

\noindent ii.) A bottom-up approach, where one tries to uncover patterns
of observed masses and mixings from what has been observed experimentally.

\noindent iii.) A more pragmatic approach, where one looks for connections
between physical phenomena.

\noindent Given the limitations of time and space I will not attempt to
be comprehensive here, but will illustrate each of these approaches by means of
some relevant examples. \cite{MS}

\subsection{Extended Symmetries}

Interesting insights on neutrinos emerge in Grand Unified Theories
(GUTs), \cite{Langacker} where quarks and leptons are in the same
multiplet(s). In the simplest GUT $SU(5)$ the right-handed neutrino
transforms as a singlet under the group. Using that $\psi^c_R$
transforms as a left-handed state, in $SU(5)$ the left-handed quark and
lepton multiplets are organized in the following multiplets:
\begin{equation}
  10=\{Q_L, ~u^c_R, ~\ell^c_R\}~~; ~~\bar{5}=\{d^c_R,~ L_L\}~~;~~
1=\{\nu^c_R\}.
\end{equation}
 As a result, because $\nu_R$ is a singlet, there are no constraints on
the mass term $m_S$.
Hence, there is no direct connection between the scale of
$SU(5)$-breaking, $M_{\rm{GUT}}$, and that of (B-L)-breaking, given by $
m_S$.

Matters are different in $SO(10)$, where all quarks and leptons,
including $\nu_R$, are in the same representation:
\begin{equation}
 16=\{Q_L,~ u^c_R,~ d^c_R,~ L_L, ~\ell^c_R,~\nu^c_R\}.
\end{equation}
Because $U(1)_{\rm {B-L}}$ is contained in $SO(10)$, for this GUT now
$U(1)_{\rm{B-L}}$ is a local, not a global, symmetry. Thus the singlet
mass $m_S$ of right-handed neutrinos is naturally 
related to the GUT breaking scale,  if (B-L) is broken at that scale.
However, in $SO(10)$ there can be an initial symmetry breaking of the GUT group which
preserves (B-L), with this symmetry then being broken at an intermediate
scale. An example is provided by the Pati-Salam \cite{PS} symmetry breakdown chain
$ SO(10)  \to SU(3)\times SU(2)_L \times SU(2)_R \times U(1)_{\rm{B-L}}
\to SU(3)\times SU(2)\times U(1).$
 Hence, in $SO(10)$ models it is possible to contemplate a range of
masses for $m_S$, ranging from perhaps $10^{10 }$ GeV  to  $10^{16}$
GeV. 

Because  the product of two 16-dimensional representations in $SO(10)$
contains a 10, a symmetric 126, and an antisymmetric 120
representations, one can consider Yukawa interactions containing
$SO(10)$ Higgs fields in these representations. In particular the Higgs
fields transforming according to the 126 representation, $126_H$, contains
a SM singlet field carrying (B-L) charge, so that naturally $m_S \sim <
126_H>$. The simplest $SO(10)$ model contains just a $10_H$ and a
$126_H$ and, using either $U(1)_{\rm{PQ}}$ or supersymmetry to eliminate
any $\overline{10_H}$ couplings, \cite{BM} has just two Yukawa coupling terms
leading to quite a restrictive mass spectrum, so that 
\begin{equation}
L_{\rm{Yukawa}}= \Gamma_D 16.16. 10_H + \Gamma_S 16. 16.
\overline{126_H},
\end{equation}
with $m_D = \Gamma_D <10_H> ~<<~ m_S = \Gamma_S <\overline{126_H}>$.

A different class of $SO(10)$ models, rather than using an elementary
$126_H$, replaces this Higgs field by a composite Higgs term made up of low-
dimensional fields: $126_H  \to (16_H *16_H) / M$. \cite{composite} Then,
even if $U(1)_{\rm{B-L}}$ is broken at the GUT scale, $m_S$ is at an
intermediate scale: $ m_S \sim  M^2_{\rm{GUT }}/ M$. Irrespective of
whether $126_H$ (and other Higgs) are composite or not, to obtain a
realistic spectrum and mixings for the charged fermions and  neutrinos,
in general one imposes some additional flavor symmetry on the models -
typically $U(1)\times D$, where $D$ is some discrete symmetry. The
remnants of these symmetries should then be visible in the mass
matrices. Some examples of $SO(10)$ models which produce realistic
quark, charged lepton, and neutrino masses are discussed by Mohapatra
\cite{Mohapatra} in this meeting.

\subsection{Patterns in the Neutrino Sector}

A different approach (more bottom-up) is to try to divine from the data
on neutrino masses and mixings the structure of the underlying theory.
In a 3 neutrino framework, ignoring the LSND \cite{LSND} result for
now, one has \cite{PDG}
\begin{eqnarray} 
   |\Delta m^2_{32}| &= \Delta m^2_{\rm{atmos}}& = (2.4 \pm
0.3)\times10^{-3} \rm{eV}^2, \\
  | \Delta m^2_{21}|& =\Delta m^2_{\rm{solar}}&  = (7.9 \pm 0.4)
\times 10^{-5} \rm{eV}^2 .
\end{eqnarray}
 These mass squared differences are consistent  both with a hierarchical spectrum of masses, $
m_3 ~>>~ m_2~ >>~m_1$ (or an inverted hierarchy), and other
patterns (e.g. $m_3~ >>~ m_2 \simeq m_1$).

As far as mixing goes \cite{PDG}, two of the angles in the leptonic
mixing matrix $U_{\rm{PMNS}}$ are large,  $s_{23}\simeq 1/\sqrt {2};
s_{12} \simeq 1/2 $  (actually, $ s_{12}\simeq 0.56)$, while the third
is bound at $3\sigma$ by  $s_{13}< 0.22$ and so could be a small angle. Thus,
approximately, letting $s_{13}=\epsilon$, one has
\begin{equation}
U_{\rm{PMNS}} \simeq
\left[\begin{array}{ccc}\frac{\sqrt{3}}{2}&\frac{1}{2}&\epsilon e^{-i
\delta}\\ -\frac{1}{2\sqrt{2}}&\frac{\sqrt{3}}{2\sqrt{2}}&\frac{1}{\sqrt{2}}
\\ \frac{1}{2\sqrt{2}}&-\frac{\sqrt{3}}{2\sqrt{2}}&\frac{1}{\sqrt{2}}
\end{array} \right].
\end{equation}
where $\delta$ is a, not yet determined, CP-violating phase.

It is often assumed \cite{Altarelli} that the patterns seen reflects some
(perhaps approximate) family symmetry in the neutrino mass matrix
\begin{equation}
 M_{\nu} = U^*_{\rm{PMNS}} m_{\nu~ \rm{diag}} U^{\dagger}_{\rm{PMNS}}
\end{equation}
 where $ m_{\nu~ \rm{diag}}=\rm{diag}~[m_1, m_2e^{i\alpha_2},
m_3e^{i\alpha_3}]$. Here $\alpha_2$ and $\alpha_3$ are two other unknown
(so called, Majorana) CP- violating phases. Because the uncertainty in
the neutrino masses $m_i$ is large, and the Majorana phases $\alpha_i$
are unknown, it is difficult to reconstruct the matrix $M_{\nu}$. It is
better, instead,
 to focus on the mixing angles.

In this respect, an interesting starting point is provided by matrices
$M_{\nu}$ which produce $s_{13}=0$ and give maximal mixing $s_{23} =
1/\sqrt{2 }$. It is easy to see that matrices $M_{\nu}$ that are 2-3
symmetric \cite{Lam}
\begin{equation}
M_{\nu}=\left[\begin{array}{ccc}X&A&A\\A&B&C\\A&C&B \end{array}\right]
\end{equation} 
precisely accomplish this. This 2-3 permutation symmetry can be part of
a larger discrete or continuous symmetry. Many models exist, based on
the groups: $A_4$, $S_3$, $Z_4$, $D_4$. \cite{MS} Because this 2-3
symmetry is approximate in nature (after all $m_{\mu} \neq m_{\tau}$),
the way that the symmetry is broken determines the correlation among the
mixing angles. Thus, $s^2_{23} - \frac{1}{2} = F(s_{13})$, where the model dependent function $F(s_{13})$ is normalized so that $F(0)=0$.

A different bottom-up approach  pursued by Smirnov and collaborators
\cite{QLC} uses the observation that
\begin{equation}
 \theta_{12}+\theta_{12}^{\rm{had}} \simeq
\theta_{23}+\theta_{23}^{\rm{had}} \simeq \frac{\pi}{4}.
\end{equation}
  to build models that display this quark-lepton complementarity (QLC).
It may well be that the above relation is a coincidence,  but it is an interesting
avenue to pursue.

I will illustrate these models with an example due to Minakata and
Smirnov. \cite{MS2} Recall that the mixing matrices are obtained after
combined unitary transformations of the leptons and quarks. Namely, $
   U_{\rm{PMNS}} = U_{\ell}^{\dagger} U_{\nu}$, while $ U_{\rm{CKM}} =
U_u^{\dagger} U_d.$
 Imagine that mixing in the doublet Higgs sector comes only from the
"down" side (thus $U_u = 1$) and that some GUT forces $U_{\ell} = U_d  =
U_{\rm{CKM}}$. Then one  gets an interesting QLC relation if  the seesaw
mechanism  forces the neutrino unitary matrix $U_{\nu}$ to be of the bi-maximal
form:
\begin{equation}
U_{\nu}=U_{\rm{bm}}=\frac{1}{2}\left[\begin{array}{ccc}\sqrt{2}&\sqrt{2}
&0\\-1&1&\sqrt{2}\\ 1&-1&\sqrt{2} \end{array}\right].
\end{equation}
If this is so, then $U_{\rm{PMNS}}= U^{\dagger}_{\rm{CKM}} U_{\rm{bm}}$.

\subsection{Connecting Physical Phenomena}

A third approach for exploring the neutrino window for physics beyond
the SM is by looking for physical interrelations among phenomena. The
best example of this is provided by leptogenesis, \cite{FY} the
generation of the Universe's matter-antimatter asymmetry from a
primordial lepton asymmetry. The ratio of baryon to photon density in
the Universe now, $\eta = n_ B/ n_{\gamma}$, is a measure of the
primordial  matter-antimatter asymmetry in the Universe. \cite{BPY}
This ratio
is well determined by WMAP \cite{WMAP} and by Big Bang Nucleosynthesis (BBN),
\cite{Steigman}  and one finds
 $\eta =(6.097 \pm 0.206) \times 10^{-10}.$

The heavy neutrinos needed for the seesaw provide a nice mechanism for
generating $\eta$. In this leptogenesis scenario \cite{FY}  a primordial
lepton-antilepton asymmetry is generated from out of equilibrium decays
of the heavy Majorana neutrinos [$N \to \ell\Phi~;  N \to
\ell^{\dagger}\Phi^{\dagger}$]. However,
because the (B+L)-current is anomalous, \cite{t'H} sphaleron processes
\cite{KM} transmute this lepton number asymmetry into a baryon number
asymmetry. \cite{KRS} In the SM one can show \cite{HT} that about
one-third of the produced lepton asymmetry becomes a baryon asymmetry.
Thus one needs to generate $ \eta_L\simeq  2\times 10^{ -9}$ to
produce the desired value for $\eta$.

The leptonic asymmetry produced from heavy neutrino decays in
the early Universe is given by the formula \cite{BPY}
\begin{equation}
\eta_L \simeq 7 \frac{\epsilon \kappa}{g^*}.
\end{equation}
Here the factor of 7 relates the entropy density to the photon density,
while $\epsilon$, $\kappa$, and $g^*$, respectively, are
a measure of the CP-asymmetry in the decays of the heavy neutrinos, take
into account of a possible washout of the asymmetry by equilibrium
processes, and count the effective degrees of freedom at the time of
leptogenesis ($g^* \sim 100)$.
Both $\epsilon$ and $\kappa$ depend on properties of the
light neutrino spectrum.

Assuming hierarchal heavy neutrinos, $\epsilon$ is determined
by the decays of the lightest heavy neutrino, of mass $M_1$,
and is given by the formula \cite{FY}
\begin{equation}
\epsilon=-\frac{3M_1}{16\pi^2v_F^2}\frac{(\rm{ Im }~\Gamma_{\nu}^*
M_{\nu}\Gamma_{\nu}^{\dagger})_{11}}{(\Gamma_{\nu}\Gamma_{\nu}^{\dagger}
)_{11}}.
\end{equation}
Leptogenesis occurs at temperatures $T$ of $O(M_1)$ and for $T \sim
M_1\sim 10^{10}$ GeV, one needs very light neutrino masses ($m_{\nu}
\leq $ eV) to obtain the typical parameters needed for the CP asymmetry
($\epsilon \sim 10^{-6}$) and the washout ($\kappa \sim 10^{-2}$) to get
$\eta_L \sim 2 \times10^{ -9 }$.

In fact,  Buchm\"uller, Di Bari, and Pl\"umacher \cite{BDP2} have shown
that the washout factor  $\kappa$ is independent of
initial abundances (and of any pre-existing asymmetry)  if the effective neutrino mass parameter $\tilde{m_1}$
 is in the range  $10^{-3} \rm{eV} \leq \tilde{m_1} \leq 1
\rm{eV}$, precisely the range of the observed masses for neutrinos. Furthermore, $\kappa$ gets smaller as the neutrino masses increase, because
the washout rate is proportional to the sum squared of the
neutrino masses, $W\sim \Sigma_i m^2_i$. As a result, successful leptogenesis
provides an upper bound on the light neutrinos masses \cite{BDP2}
\begin{equation}
 m_i \leq  0.1 ~{\rm{eV}}.      
\end{equation}
This result is perfectly consistent with observations (excluding the
LSND results) on light neutrinos.

Nevertheless, there are some clouds on the horizon. Because $\epsilon$
cannot be to small for leptogenesis to work, this provides a lower bound
on  the mass of the  lightest heavy neutrino, $M_1$. \cite{DI} For
$\tilde{m}_1 \geq 10^{-3} \rm {eV}$, where the result for $\eta_L$ is
independent of pre-existing conditions, one finds \cite{BDP}
\begin{equation}
M_1 \geq 2\times 10^9~ {\rm{GeV}}.  
\end{equation}
However, the fact that leptogenesis occurred at temperatures $T\sim M_1>
2\times 10^9 {\rm{GeV}}$  has significant import for supersymmetric
theories. It turns out that if the reheating temperature after inflation
$T_R$ is too high one overproduces gravitinos. This has catastrophic
consequences, since the decay products produced in gravitino decays
destroy the light elements produced in Big Bang Nucleosynthesis. To
avoid troubles, one must require a much lower reheating temperature
$T_R \leq 10^7 {\rm{GeV}}$. \cite{Kawasaki} But leptogenesis argues
$T_R \geq M_1 \geq 2\times 10^9 $ GeV!
There are solutions to the gravitino problem, but these in general alter
the  "normal" SUSY expectations coming from supergravity. For example,
one needs either very heavy gravitinos ($m_{3/2} \geq $100 TeV)
\cite{heavy} or very light gravitinos ($m_{3/2} \leq $ 1 GeV), so that
the gravitino is the LSP. \cite{light} 

Lepton
flavor violation (LFV) provides another example of tension between SUSY
and leptogenesis. For example, predictions for radiative muon decays,
$\mu \to e\gamma$, in SUSY theories, although model dependent, are
sensitive to the mass of the heavy neutrinos.  As a result these LFV
processes constrain the heavy neutrino spectrum from above. \cite{Petcov} Obviously,
future experimental information will be crucial. Indeed, asking compatibility
between SUSY and leptogenesis may lead to testable experimental predictions
and insights into neutrino physics.

\section{Wild Speculations}

 \subsection{LSND Musings}

There well may be more surprises in the neutrino sector. A prime example
is provided by data  from the LSND experiment. \cite{LSND}. This
experiment reported a positive signal of $\bar{\nu}_{\mu} \to
\bar{\nu}_e$ oscillations  for neutrino mass squared differences $\Delta
m^2 \sim \rm{eV}^2$ and mixing $\rm{sin}^2 \theta\sim 10^{-3}$. This
result, obviously, 
 cannot be reconciled in a three-neutrino framework, as we already have
two other distinct mass squared differences from solar and atmospheric
neutrino oscillation experiments. We  need to await results from the
Mini BooNe  experiment \cite{mini} for confirmation, but if the LSND
phenomena is true it requires different physics than the one we have
been discussing up to now, involving sterile neutrinos or CPT
violation.

It is easy to find candidates for sterile neutrinos in string/GUT
theories. For example, if there is an underlying $E_6$  symmetry,
\cite{E6} in its 27-dimensional representation, which contains the
quarks and leptons, one has nine states which are color singlets. Among
these nine states there are two $ SU(2) \times U(1)$ singlets. One of them
is the right-handed neutrino, but there is also an additional state in each
27-dimensional multiplet with the same SM properties as the $\nu_R$.
Obviously such states, or mixtures of these states with the $\nu_R$
states, can play the role of sterile neutrinos.

 The challenge, however,
is not finding schemes where there are sterile neutrinos but to get the
mass of these states to be of order $m_{\rm{st}} \sim$ 1 eV. The normal
trick is to use some discrete symmetry to set $ m_{\rm{st}}=0$ and then
get a small mass from the breaking of this symmetry. For example,
Mohapatra {\it{et al}} \cite{M23} use a 2-3 symmetry for these
purposes. However, in my view, these models are quite complex and not
that well motivated. Furthermore, the phenomenology of, so called, 3+1
models with one additional sterile neutrino is shaky, \cite{3-1pheno}
but 3+2 models might provide a better fit. \cite{3-2pheno}.

Invoking CPT violation to account for the LSND result is a much bolder
suggestion. \cite{Barenboim} If CPT is broken in the neutrino sector,
one  expects differences between $\nu_{\mu} \to \nu_e$   and $
\bar{\nu}_{\mu} \to \bar{\nu}_e$  oscillations. In particular, there is
no reason why the mass squared differences between particles and
antiparticles be the same. So, one can fit the LSND result in a three
neutrino context, since there are now 4 possible mass squared
differences. Unfortunately, as  Gonzales- Garcia, Maltoni and Schwetz
showed, \cite{GGMS} the phenomenology does not work better in this case
either. Keeping all mass squared differences free, a global fit of all
data except that of LSND is in agreement with the CPT conserving
solution $\Delta m^2=\Delta \overline{m}^2$. Thus only the LSND result
requires CPT violation. However, a recent paper by Barenboim and
Mavromatos \cite{BaMa} claims that matters are improved by including
quantum decoherence effects. This is not unexpected, since these effects
essentially add more parameters to the fit.  Obviously, more data is
badly needed!

\subsection{Mass Varying Neutrinos}

Recently, Fardon,
Nelson and Wiener (FNW) \cite{FNW} suggested that neutrinos may have
environment-dependent masses which may play a role in helping to explain
the dark energy in the Universe. The existence of dark energy, which
acts as a fluid with negative pressure, was first inferred from
supernova data indicating that the Universe's expansion was
accelerating. \cite{acc} This result has now been confirmed by data from
WMAP. \cite{WMAP} If one combines
WMAP data with that from the Supernova Legacy Survey \cite{SLS} one can
determine the dark energy
 equation of state $w=p/\rho$
and one finds a result 
which is consistent with the dark energy being a cosmological constant
($w_{\Lambda} = -1$).

Even though neutrinos are a subdominant component of the Universe's
energy density now, contributing less than 1.5\%, what Fardon {\it{et
al}} suggested is that $ \rho_{\rm{dark~energy}}$
tracks the energy density in neutrinos, $\rho_{\nu}$. More precisely, in
their picture neutrinos and dark energy are assumed to be coupled. In
the non relativistic regime they examined, the total dark sector energy
density has two components. A neutrino piece and a dark energy piece
which is coupled to the neutrinos since  $ \rho_{\rm{dark~energy}}$
depends on the mass of the neutrinos. Considering for simplicity just
one generation of neutrinos, one has
\begin{equation}
 \rho_{\rm{dark}} =  m_{\nu}n_{\nu}+  \rho_{\rm{dark~energy}}(m_{\nu}),
\end{equation}
where $n_{\nu}$ is the neutrino density.
Because the two components are coupled, neutrino masses are determined
dynamically by minimizing the above equation. As a result, in the FNW  model, neutrino masses depend on
density, $m_{\nu}= m_{\nu}(n_{\nu})$.

The equation of state for the dark sector follows from the energy
conservation equation
and in NR limit one finds \cite{FNW}
\begin{equation}
   w +1= \frac{ m_{\nu} n_{\nu}} {\rho_{\rm{dark}}} = \frac{ m_{\nu}
n_{\nu}} { m_{\nu} n_{\nu } +\rho_{\rm{dark~energy}}}. 
\end{equation}
If $w\simeq -1$ the neutrino contribution to $\rho_{\rm{dark }}$ is a
small fraction of $\rho_{\rm{dark~energy}}$. Furthermore, one can show
that, in its simplest version, the FNW model is
equivalent to having a cosmological constant, which runs with the
neutrino mass. \cite{Peccei} Thus
$\rho_{\rm{dark~energy}} =  -p_{\rm{dark~energy}}\equiv V(m_{\nu})$
 and $m_{\nu}= m_{\nu}(n_{\nu})$ is determined by the minimization
condition \footnote{This equation is written still in the NR limit, but
it can be generalized to neutrinos of arbitrary velocity. \cite{Peccei}}
\begin{equation}
   n_{\nu}+ \frac{\partial V(m_{\nu})}{ \partial m_{\nu}}  = 0
\end{equation}

A typical example of a potential which satisfies the above equation is
$V(m_{\nu}) ~\sim m_{\nu}^{-\alpha}$. \cite{Peccei} In this case, $
\alpha$ is determined by the value of $w$ now, $\alpha= - (1 +w_o)/w_o$
and the neutrino mass drops rapidly with increasing density, so that
neutrinos of mass $m^o_{\nu}= 3$ eV now are relativistic ($m_{\nu}=T$)
already at temperatures of order $T \simeq 3 \times 10^{-3}$  eV.  For
$w_o=-0.9$, the equation of state has $w\simeq w_o$  up to a redshift of
order $z\simeq 10$, but by $z=20$ one has $w\simeq -0.25$. At high $z$,
eventually $w \to 1/3$, as the dark energy density reduces to that of
relativistic neutrinos.

There are many issues one can raise concerning neutrino models of dark
energy. For instance, what physics fixes the running cosmological
constant potential $V(m_{\nu})$ ? Or,
what dynamical principle demands that $\rho_{\rm{dark}}$ be stationary
with respect to variations in the neutrino mass?  In addition, the
simplest version of mass varying neutrino models described above leads
to a dynamical instability, since the speed of sound squared,
$c_s^2=\partial p/\partial \rho$, is negative in the late stages of the
Universe's evolution. \cite{sound} \footnote{This problem can be avoided
in more elaborate models. \cite{avoid}}    
Nevertheless,  the idea of variable mass neutrinos is intriguing since
it associates the dark energy sector, through the seesaw mechanism, to
the $SU(3)\times SU(2)\times U(1)$ singlet sector connected with heavy
neutrinos. Because this sector is difficult to probe, it is easy to imagine
that the physics which determines the Universe's late dynamics lurks
there.

\section{Concluding Remarks}

Neutrino physics has opened windows into phenomena beyond the SM,
associated in the first instance with (B-L)-breaking. In my view, the
smallness of the masses of neutrinos relative to those of quarks and
leptons is a reflection of the hierarchy between the scale of (B-L)
breaking and the Fermi scale $v_F$ embodied in the seesaw mechanism.
\cite{seesaw} Attempts to understand the details of neutrino masses and
mixings have provided hints of possible flavor symmetries and of
unification, although no unequivocal theoretical direction has yet
surfaced. Deeper puzzles and mysteries may well surface in the future,
involving neutrinos and their role in the Universe. We await more
experimental data, particularly connected with CP-violation in the
neutrino sector and on the behavior of dark energy with increasing
redshift.

\section*{Acknowledgments}

This work was supported in part by the Department of Energy under Contract No. FG03-91ER40662, Task C.

\end{document}